\documentclass[%
 reprint,
 amsmath,amssymb,
 aps,
prstab,
]{revtex4-1}

\usepackage{graphicx}
\usepackage{dcolumn}
\usepackage{bm}
\usepackage{gensymb}

\begin{document}


\title{Online storage ring optimization using dimension-reduction and genetic algorithms}

\author{W. F. Bergan}
\email[]{wfb59@cornell.edu}
\author{I. V. Bazarov}
\author{C. J. R. Duncan}
\author{D. B. Liarte}
\author{D. L. Rubin}
\author{J. P. Sethna}
\affiliation{Cornell University, Ithaca, New York 14853, USA}

\begin{abstract}
Particle storage rings are a rich application domain for online 
optimization algorithms. The Cornell Electron Storage Ring (CESR) has 
hundreds of independently powered magnets, making it a high-dimensional test-problem for algorithmic tuning. We investigate 
algorithms that restrict the search space to a small number of linear 
combinations of parameters (``knobs'') which contain most of the 
effect on our chosen objective (the vertical emittance), thus enabling 
efficient tuning. We report experimental tests at CESR that use 
dimension-reduction techniques to transform an 81-dimensional space to 
an 8-dimensional one which may be efficiently minimized using one-dimensional parameter scans. We also report an experimental test of a 
multi-objective genetic algorithm using these knobs that results in 
emittance improvements comparable to state-of-the-art algorithms, but 
with increased control over orbit errors.

\end{abstract}

\maketitle

\section{Introduction}

Despite the great care taken in accelerator design and fabrication, inevitable magnet 
misalignments, calibration errors, and drifts will result in sub-optimal beam properties. Since the exact nature of these errors will 
not be known in advance, it is necessary 
to correct them using online techniques, i.e., operating 
directly on the real machine. Time spent tuning the machine is time 
that it is unavailable for its intended use, and so it is desirable 
that any correction procedure be fast, especially in the case of 
multipurpose facilities. Light sources and future colliders, such as 
the International Linear Collider (ILC) \cite{cite:ILC}, have hundreds 
to thousands of magnets, so an optimizer must be able to search these 
high-dimensional spaces in a reasonable time. To make tuning of such 
problems feasible, low-dimensional approximate models can be combined 
with empirical data to speed up online optimization. This paper 
reports the results of testing the performance of candidate algorithms 
in both experiment and simulation on the Cornell Electron Storage Ring 
(CESR). These results are of interest to optimal control theorists, 
demonstrating real-world success with a high dimensional test case, 
and to the accelerator community, as a working solution to a problem 
of ever greater practical importance.

An ongoing project at CESR is to determine efficient ways to minimize 
vertical emittance. Vertical emittance, being important for 
accelerator performance and sensitive to global magnet misalignments 
and strength errors, is an apt metric for evaluating online 
optimization algorithms. The method for tuning vertical emittance now 
deployed at CESR, described in \cite{cite:shanks, cite:shanks_paper}, 
is to measure causes of vertical emittance, such as the coupling and 
vertical dispersion, and then to make corrections using Levenberg-Marquardt least-squares minimization. The reach of this method is 
limited by the finite resolution of CESR's beam position monitors 
(BPMs) and therefore leaves residual vertical emittance, which is 
discernible in our high-resolution beam size measurements. Independent component 
analysis (ICA) \cite{cite:ICA}, dispersion-free-steering (DFS) 
\cite{cite:DFS}, and a low-emittance tuning (LET) algorithm 
\cite{cite:let_algorithm} have also been tried at other accelerators, 
but these too suffer from reliance on accurate dispersion measurements 
for proper operation. Scanning magnet settings to tune directly on 
vertical emittance can yield further improvements. Researchers at the 
Swiss Light Source (SLS) have had success in reducing vertical emittance 
by varying the strengths of useful correctors randomly and observing 
the resulting emittance \cite{cite:sls}. Unguided searches are time-inefficient, and Huang et al. have improved upon this method by 
introducing the robust conjugate direction search (RCDS) algorithm 
\cite{cite:huang_rcds1,cite:huang_rcds2}.

The RCDS method makes use of simulation to obtain the Hessian matrix 
for the merit function with respect to corrector magnets and makes 
corrections to the real machine using the eigenvectors of this matrix. 
These eigenvectors are conjugate directions, and have the property that optimizing along one 
direction does not require retuning of another direction, insofar as 
the simulation is correct and nonlinearities in the merit function are 
small. It then makes one-dimensional scans of each search direction 
and uses a quadratic fit to determine the minimum. As the algorithm 
moves in the space of machine states, it adjusts the search directions 
based on acquired data. The RCDS method requires tuning a number of 
knobs equal to the number of independently tunable parameters, and so the time to 
execute a full optics correction grows linearly with the number of 
available independent magnet groups.

Researchers at SLAC applied genetic algorithms to the SPEAR3 storage 
ring \cite{cite:huang_genetic,cite:tian_genetic} and judged 
performance to be unacceptable because of the time for convergence and 
influence of measurement errors. However, such algorithms have been 
successfully applied to the nine-dimensional problem of optimizing the 
beam transmission through one of the GSI beamlines 
\cite{cite:fair_optimizer}.

High-dimensional models are {\em sloppy} if their predictions are accurately captured by low dimensional approximations. It is an ongoing research program to investigate the common features of sloppy models that explain why, and for which set of problems, dimension reduction is successful \cite{cite:sethna0}. This research motivates the application of dimension-reduction techniques to the problem of accelerator tuning. Evidence suggesting the effectiveness of such techniques includes the observation by Marin et al. that in the course of designing
correction schemes for the final focus system for the Compact Linear
Collider only a handful of the knobs proved useful for
corrections \cite{cite:marin}, although no special emphasis was placed
on this phenomenon at the time. These ideas also appear in orbit-correction techniques, where one often uses only the first few singular vectors of the singular value decomposition (SVD) when constructing the pseudoinverse of the orbit response matrix in order to filter out noise or avoid singular behavior \cite{cite:loco, cite:mia, cite:orbit_svd}. As a tool, dimension reduction promises to widen the scope of optimization algorithms that are feasible for high-dimensional systems.

We use an SVD to extract 8 effective knobs from the 81-dimensional space of useful corrector magnets at CESR and test the utility of these new knobs as part of two very different tuning algorithms. Running the RCDS algorithm with this reduced set of knobs, we obtain beam sizes comparable to what we obtain with our standard tuning based on direct measurement and correction of dispersion and coupling. We also use the knobs as the genes in a multi-objective genetic algorithm aiming to fix both the vertical beam size and the orbit near the narrow-aperture undulators. Although there are more convenient techniques for minimizing both objectives, our purpose is to develop a technique applicable to the wider range of optimization problems that arise in accelerator operation, and it is useful to test new algorithms in regions where proven methods already exist. We find in simulation that such algorithms show improved rates of convergence when varying the 8 effective knobs found by the SVD, as opposed to the 81 raw magnet values. In experiment, we obtain beam sizes comparable to the results of directly measuring and correcting the dispersion and coupling.

Our approach differs significantly from the use of SVDs in orbit-correction: in the latter case, one has a Jacobian matrix and signed measurements, so it is possible to determine how far and which way to turn the knobs without further measurements. For that case, the primary use of the SVD is to construct the pseudoinverse and avoid issues arising from noise and over-constrained or under-constrained problems. In our case, we are minimizing a positive-definite scalar (the emittance), and so have no directional information from the Hessian matrix. We must instead do a search of parameter space with many intermediate measurements, in which case it is very useful to reduce the dimensionality of the space. Our approach is complementary to ICA and other techniques making use of auxiliary measurements, since our method does not rely on those measurements and so may be applied in cases where one has reached the limit of their accuracy or even when such diagnostics are lacking altogether.

The structure of the paper is as follows. In Section 
\ref{sec:sloppy_models}, we will introduce sloppy models and their relation to dimension-reduction techniques. In Section \ref{sec:cesr}, we will 
provide an overview of the layout of CESR. In Section 
\ref{sec:single_obj}, we will apply the sloppy-model-based dimension-reduction to single-objective tuning of the vertical emittance in both 
simulation and experiment. In Section \ref{sec:multi_obj}, we will 
discuss multi-objective genetic algorithms and show the results of 
applying them along with the knobs found by sloppy-model-based dimension-reduction to 
minimize beam size and orbit errors in both simulation and experiment.
In Section \ref{sec:conclusion}, we summarize our results and present future directions.

\section{Sloppy Models}\label{sec:sloppy_models}

In using simulation to guide online tuning of the real machine, we 
reduce the dimensions of the accelerator parameter space, motivated by 
the concept of sloppy models, which has been successfully applied to a 
large variety of systems 
\cite{cite:sethna1,cite:sethna2,cite:sethna3,cite:sloppy_chapter,cite:transtrum}. This concept posits 
that the system behavior is effectively described by a set of 
``eigenparameters,'' combinations (typically, non-intuitive) of 
original control parameters, and is in many ways similar to principal 
component analysis \cite{cite:PCA}. These eigenparameters, or knobs, 
have the useful property that when ordered in terms of the size of 
their effect on the objective, subsequent eigenparameters are 
exponentially less important than prior ones. Using only the first few 
so-called ``stiff'' eigenparameters (corresponding to larger 
eigenvalues) to describe the system therefore retains most of the 
information contained in all the parameters. There is generally no sharp cutoff between the more and less useful eigenparameters, so that the choice of how many to actually use 
is somewhat arbitrary.

Although sloppy systems and underdetermined systems are often related, neither is a subset of the other; sloppy systems may exist where the number of measurements used to characterize the system is greater than the number of parameters in the model, as shown in the Robertson model in \cite{cite:sloppy_chapter}, and having an underdetermined system only implies a zero eigenvalue for each parameter greater than the number of measurements, but does not necessarily imply that the non-zero eigenvalues decrease exponentially in importance. In addition to the linear case discussed in this paper, research in the area of sloppy models has also been effectively applied in highly nonlinear systems. Researchers have noted that in many systems certain parameter combinations can be taken to infinity with little alteration of the model predictions, permitting a reduction of dimensions even in nonlinear cases \cite{cite:transtrum}.

To obtain the eigenparameters, one may construct the Hessian matrix 
to express the second derivatives of the objective with respect to all 
pairs of parameters, and then take the SVD to obtain 
the eigenvalues and eigenvectors
\footnote{The SVD has been shown to be more numerically stable than an 
eigendecomposition \cite{cite:sethna1}, although, for the most 
important eigenvectors, the differences are negligible.}. A very similar procedure is used for the generation of the RCDS knobs, and so the knobs we obtain are also conjugate directions. Our main focus, however, is its ability to reduce the number of relevant dimensions of the problem, permitting tuning in what would otherwise be an infeasibly large search space.

\begin{figure}[!htbp]
\begin{center}
\includegraphics[width=1.0\columnwidth]{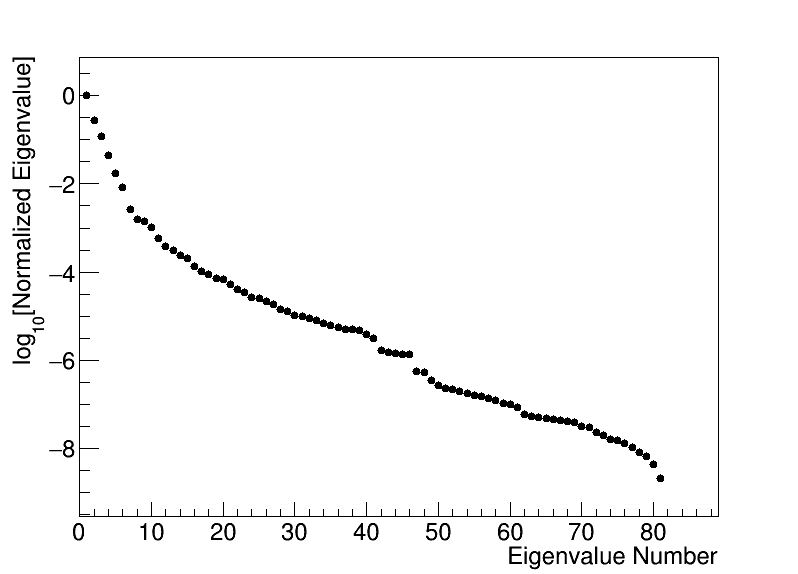}
\end{center}
\caption{\label{fig:sing_val} The spectrum of eigenvalues obtained 
from the Hessian matrix of the emittance, normalized so that the first 
eigenvalue is equal to one. Note the logarithmic vertical scale. The Hessian 
matrix and its spectrum are available as supplementary material.}
\end{figure}

In our case, 
we computed the Hessian matrix for the vertical emittance from our BMAD 
model \cite{cite:bmad} of CESR in the 81-dimensional space of control 
parameters that comprise CESR's 57 vertical kickers and 24 skew 
quadrupoles, since these were the magnets which had significant effects 
on the vertical emittance \footnote{We had found that we obtained better 
eigenparameters for fixing the vertical emittance when we did not use the 
Hessian for the ideal lattice, but rather one which had realistic magnet 
misalignments and corrections according to our standard procedure for minimizing the dispersion and coupling and that the eigenparameters obtained from the Hessians of two 
different misaligned lattices are more similar to one another than to 
those obtained from the Hessian of the ideal lattice.}. Expressing the 
vertical emittance to second order near its minimum, we have 
$\epsilon(\vec{x}) \approx \epsilon (\vec{x}_0) + \dfrac{1}{2} 
(\vec{x}-\vec{x_0})^TH(\vec{x}-\vec{x_0})$, where $\vec{x}$ is a 
vector of corrector magnet strengths, $\vec{x}_0$ is the vector of 
corrector magnet strengths which gives us the minimum vertical 
emittance, $\epsilon(\vec{x})$ is the vertical emittance given some 
corrector magnet strengths, and $H$ is the Hessian matrix. If $H$ has 
normalized eigenvectors $\vec{h}_i$ and eigenvalues $\epsilon_i$, we 
may express the emittance as $\epsilon(\vec{x}) \approx \epsilon 
(\vec{x}_0) + \dfrac{1}{2} \sum_i \epsilon_i (\vec{h}_i \cdot 
(\vec{x}-\vec{x_0}))^2$. Improvements obtained using the $i^{th}$ 
eigenvector are therefore proportional to the corresponding 
eigenvalue, assuming that all eigenvectors are displaced by the same 
amount from their optimal values. In the presence of noise in the emittance measurement of characteristic size $\sigma$, and if one expects the magnet strengths to require tuning of amplitude $y$, the above equation implies that knobs with corresponding $\epsilon_i < 2 \sigma/y^2$ will not provide any visible emittance improvement. As is shown in Fig. 
\ref{fig:sing_val}, the eigenvalues decay exponentially, confirming 
the sloppiness assumption.

\begin{figure}[!htbp]
\begin{center}
\includegraphics[width=1.0\columnwidth]{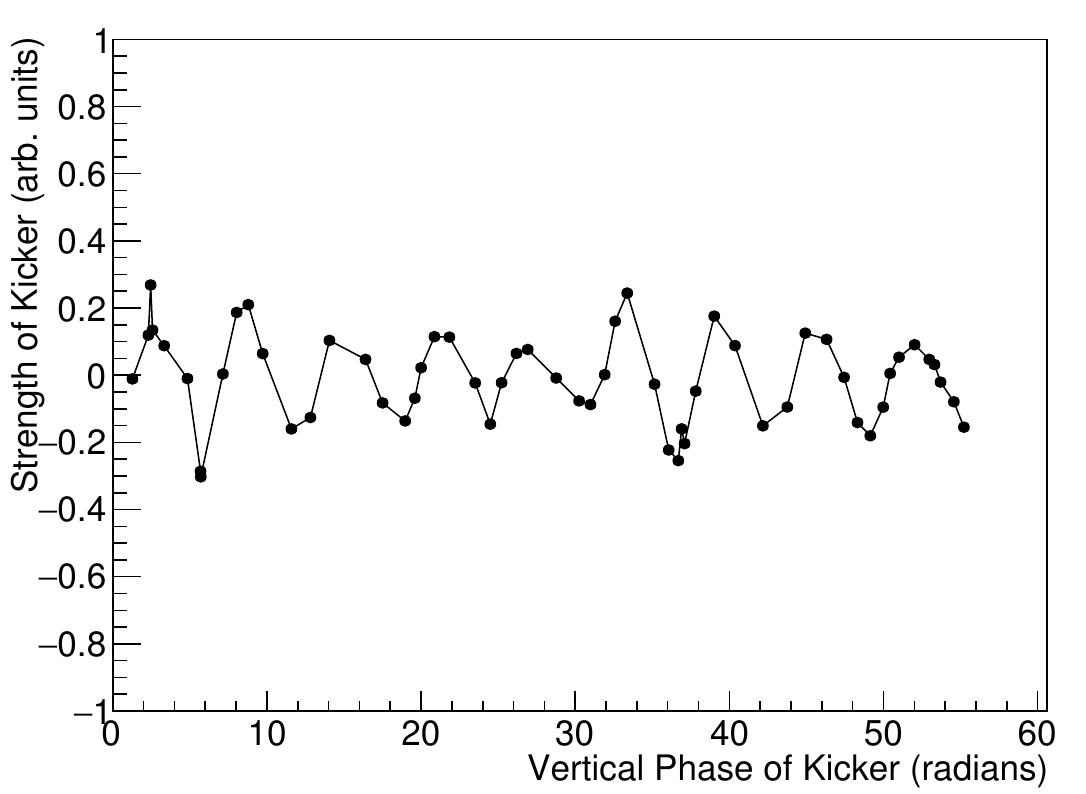}
\end{center}
\caption{\label{fig:knob_shape} The relative strengths of the kickers used in the first eigenvector as a function of their vertical betatron phase in CESR. Their arrangement in the shape of a betatron wave is readily apparent. The vertical tune is 8.79, nearly matching the 9 crests of the wave.}
\end{figure}

It is also interesting to study the magnets used in the eigenvectors to see if there is any underlying structure. Fig. \ref{fig:knob_shape} shows the relative strengths of the 57 vertical kickers used in the first eigenvector as a function of their vertical betatron phase in the storage ring. They have a periodic structure with a total number of crests (9) close to the vertical betatron tune of 8.79, and, in general, the first several knobs have the appearance of a global wave when plotted in this way. This would make sense, since a coordinated shift in kickers with the same betatron phase should have the strongest effect on machine optics. Knobs with smaller eigenvalues have more varied structures, with very high-frequency and/or highly localized components. The skew quadrupoles did not have large components in the first several knobs, although we believe that this is at least partially due to the different scales of the magnet strengths. Corresponding plots for all eigenvectors are available as supplemental material.

\section{CESR Overview}\label{sec:cesr}

CESR is a storage ring that operates at 5.3 GeV with counter-rotating 
electrons and positrons as a light source, and at lower energies as a 
testbed for future accelerators. For our experiments, we use the 
multibunch, two-beam lattice typical of CESR's operation as a light 
source \cite{cite:chess_lattice}. The design horizontal and vertical 
emittance in this lattice are 97 nm-rad and 37 fm-rad, 
respectively, although, due to inevitable magnet misalignments, in practice we 
measure vertical emittances of roughly 20 pm-rad. All of CESR's magnets 
are independently powered, giving us flexibility in how we apply 
corrections. Additionally, it has approximately 100 beam position 
monitors (BPMs) distributed about its circumference, enabling 
measurements of orbit, dispersion, coupling, and other optics 
functions. It also has various beam size monitors, although our 
studies have been carried out exclusively with the visible synchrotron 
light beam size monitor (VBSM) \cite{cite:vbsm}. Parameters of this 
ring are shown in Tab. \ref{tab:cesr}.

\begin{table}[!hbt]
   \centering
   \caption{CESR Parameters.}
   \begin{tabular}{lc}
       \toprule
       Circumference & 768 m\\
       Energy & 5.3 GeV\\
       Horizontal Emittance & 97 nm\\
       Vertical Emittance (Ideal) & 37 fm\\
       Vertical Emittance (Actual) & 20 pm\\
       Horizontal Tune & 11.29\\
       Vertical Tune & 8.79\\
       Horizontal Beta Function at VBSM & 4.8 m\\       
       Vertical Beta Function at VBSM & 15.5 m\\
       Fractional Energy Spread & 6.5 $\times$ 10$^{-4}$\\
       \toprule 
   \end{tabular}
   \label{tab:cesr}
\end{table}

For all experimental tests, we choose beam parameters that maximize 
the sensitivity of diagnostic instruments and minimize the influence 
of collective effects, storing a single bunch with a modest current of 0.75 mA. To eliminate spurious sources of emittance, we disable the 
electrostatic separators and multibunch feedback, and attenuate the 
feedback kicker amplifier \cite{cite:shanks, cite:shanks_paper}. 

\section{Single-objective Tuning}\label{sec:single_obj}

\subsection{Simulation}\label{subsec:single_obj_sim}

\begin{figure}[!htbp]
\begin{center}
\includegraphics[width=1.0\columnwidth]{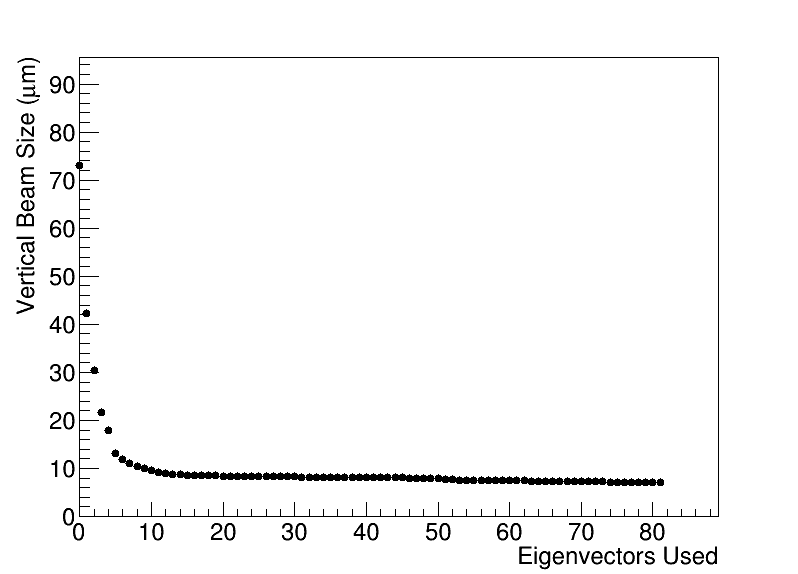}
\end{center}
\caption{\label{fig:uncorr_emit} The average simulated beam 
size over 1,000 instances of a lattice with random misalignments after 
minimizing the beam size using our first N eigenparameters. The 
average initial beam size is shown by the point lying on the vertical 
axis. Note the rapid decrease in beam size from the first few 
eigenparameters.}
\end{figure}

\begin{figure}[!htbp]
\begin{center}
\includegraphics[width=1.0\columnwidth]{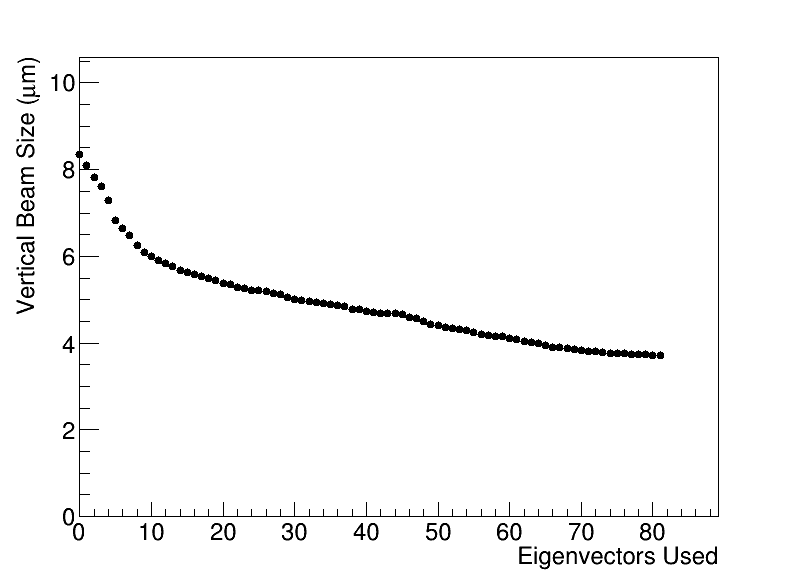}
\end{center}
\caption{\label{fig:corr_emit} The average simulated beam 
size over 1,000 instances of a lattice with random misalignments after correcting dispersion and coupling using our Levenberg-Marquardt-based tuning and then minimizing the beam size using our first N eigenparameters. The 
average initial beam size is shown by the point lying on the vertical 
axis. Although not as stark as when starting from an uncorrected lattice, we still see that the first few eigenvectors contribute a disproportionate amount to the reduction in beam size.}
\end{figure}

To test the stiffness of our knobs in simulation, we generate 1000 configurations of the ring guide field with random magnet misalignments and strength errors 
consistent with our measurement tolerances. Details of the assumed 
magnet errors may be found in the appendix of 
\cite{cite:shanks_paper}. The knob-based tuning starts with the most 
important eigenvector and makes a one-dimensional search to find the 
value of that knob which minimizes the vertical beam size at the 
location of the electron VBSM. These steps are repeated for all 81 
eigenparameters, with the results shown in Fig. 
\ref{fig:uncorr_emit}. We clearly see that almost all 
the improvement in the beam size is due to the tuning of the first few 
knobs, as we would expect for a sloppy system. Repeating this procedure, but after first using our usual Levenberg-Marquardt-based minimization of dispersion and coupling, gives the results shown in Fig. \ref{fig:corr_emit}. We again see that the first few knobs are disproportionately effective at fixing the beam size, although the contrast is not as stark as in the uncorrected case due to the fact that our Levenberg-Marquardt-based minimization makes corrections along some of the knob directions, thus reducing their utility.

We measure vertical beam size as a proxy for vertical emittance. Although 
vertical beam size also depends on the vertical beta function, 
dispersion, and coupling at the VBSM source point, our knobs do not 
change the beta function significantly (below one part per thousand), 
and the vertical dispersion and coupling are ideally zero. Therefore, while the beam size is not a perfect analogue for the emittance, it is still an interesting parameter to minimize for testing our methods.

\subsection{Experiment}

For the experimental tests, we restrict our search space to the 
leading eight eigenparameters obtained from the simulated Hessian. We 
mitigate the effects of measurement uncertainty by taking point 
averages of multiple measurements \footnote{To avoid the influence of an observed high-side tail, we make 10 measurements and take the average of those that are 
within 4 $\mu$m of the minimum value obtained for that particular set 
of measurements. If insufficient beam size measurements fit that 
criterion, we make a fresh set of ten measurements. This method allows us to 
obtain a more consistent beam size measurement and also lets us 
compute an associated error.}. We scan each knob one-by-one as ordered 
by eigenvalue using a modified RCDS algorithm. Step sizes were set to 
be 2/7 of the knob value which increased the vertical emittance by 15 
pm in simulation. This value was chosen so that when performing online 
tuning we would expect to be sure of bounding the minimum beam size 
within a reasonable number of measurements and have enough data to 
perform a decent fit. The scan ranges up to five standard deviations 
about the minimum vertical beam size. The knob setting for minimum 
vertical beam size is estimated from a quadratic fit to the data 
obtained from the scan \footnote{Based on the discussion in Section \ref{sec:sloppy_models}, the emittance should be quadratic in the values of the tuning knobs. We have verified experimentally that the beam size also has a similar quadratic dependence.}. The orbit is altered by our use of vertical correctors, and no attempt is made to hold it fixed. Each one-dimensional scan may be completed in 
a few minutes. We perform our tests starting from both an ``uncorrected'' and a ``corrected'' lattice. The difference between these two cases is that, for the latter, we have first applied our standard Levenberg-Marquardt-based tuning of dispersion and coupling. The uncorrected lattice retains historical magnet adjustments made for CESR's light-source operations and general machine usability, but no special effort had been made to minimize the vertical emittance.

\begin{figure}[!htbp]
\begin{center}
\includegraphics[width=1.0\columnwidth]{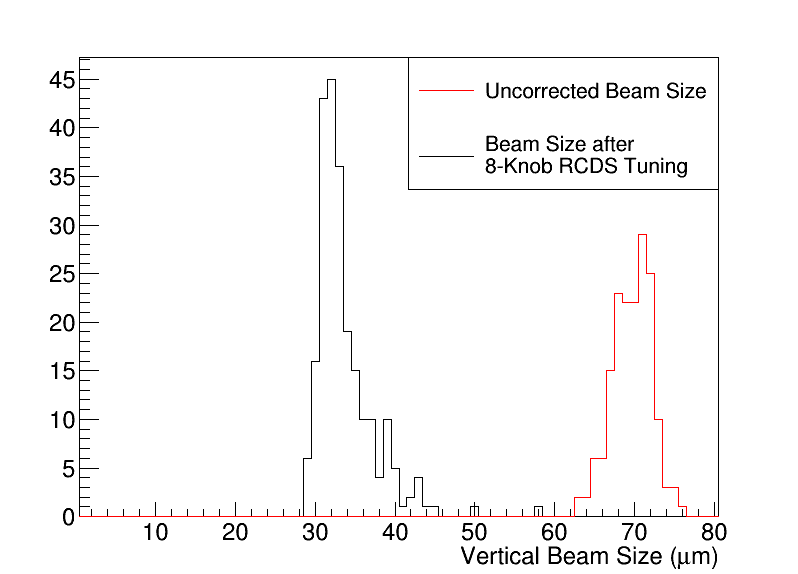}
\end{center}
\caption{\label{fig:start_uncorrected} Histogram of hundreds of 
experimental vertical beam size measurements both using the standard 
lattice for light-source operations with no additional tuning and 
after tuning with the RCDS algorithm using the best 8 eigenparameters. 
A clear improvement is observed. Note also the high-side tail of the 
low-beam-size distribution.}
\end{figure}

\begin{figure}[!htbp]
\begin{center}
\includegraphics[width=1.0\columnwidth]{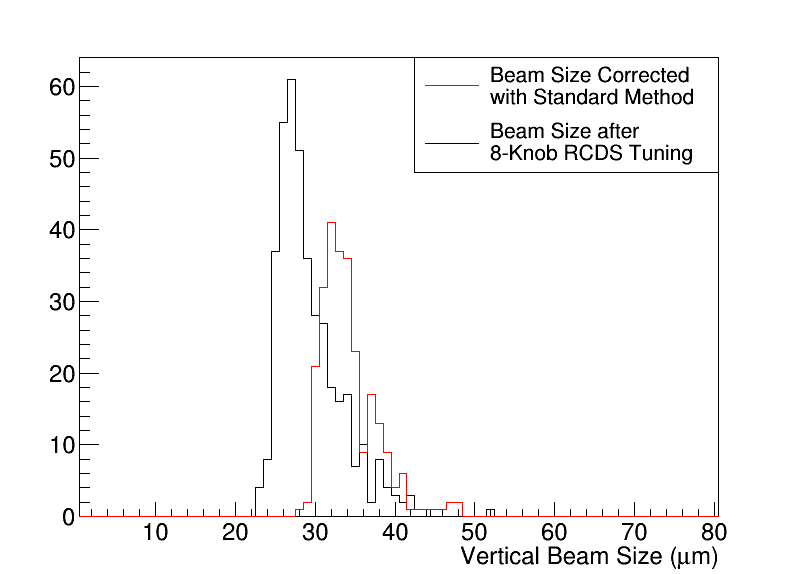}
\end{center}
\caption{\label{fig:start_corrected} Histogram of hundreds of 
experimental vertical beam size measurements both after applying just 
the standard Levenberg-Marquardt minimization of dispersion and coupling and after additional 
tuning with the RCDS algorithm using the best 8 eigenparameters. A 
modest but unambiguous improvement is observed. Note also the high-side tails of the measurement distributions.}
\end{figure}

When applying RCDS tuning to both an ucorrected and corrected CESR lattice, we obtain the 
results shown in Fig. \ref{fig:start_uncorrected} and Fig. 
\ref{fig:start_corrected}, respectively. We see that, when starting 
from an uncorrected lattice, we are able to reach beam sizes 
comparable to what the standard low-emittance tuning algorithm is able 
to deliver. When starting from a lattice already corrected by the 
standard tuning methods, we are able to bring about a clear further 
reduction in beam size.

\begin{figure}[!htbp]
\begin{center}
\includegraphics[width=1.0\columnwidth]{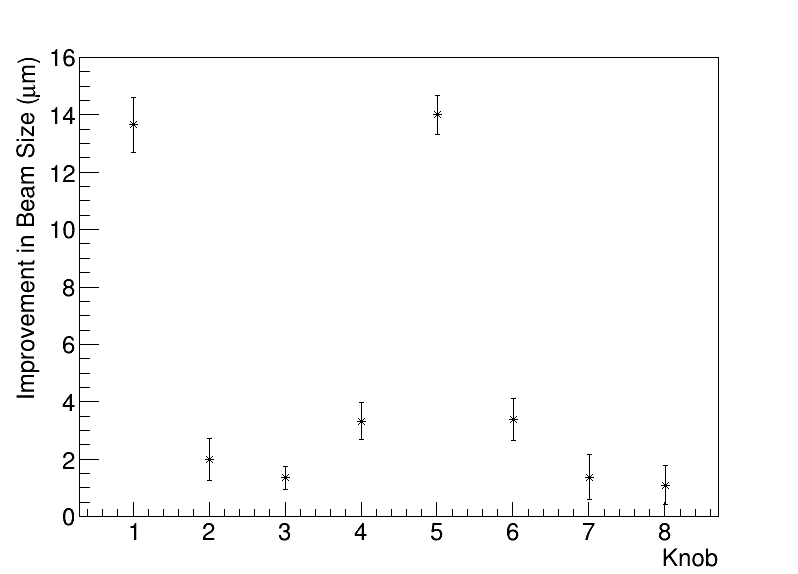}
\end{center}
\caption{\label{fig:improvement_exp} The improvement due to each knob when tuning the real machine starting from a lattice without additional corrections. Knob 5 contributes a surprisingly large amount. However, we note that it needed to be turned more than twice as far as any other knob, and more than four times as far as the first knob, which suggests that its utility stems from the fact that, when tuning the lattice for use in light-source operations, the first few knobs had already been used, and so the fifth knob is the strongest knob to have not been well-tuned. Note that, when comparing to Fig. \ref{fig:uncorr_emit} or Fig. \ref{fig:corr_emit}, the latter two plots show the beam size after tuning N knobs while this plot shows the change in beam size from tuning the N$^{\textrm{th}}$ knob.}
\end{figure}

We also examine how much each knob contributes to the improvement of the beam size, as is shown in Fig. \ref{fig:improvement_exp} for the case when starting from an uncorrected lattice. We see that the first and fifth eigenvectors are the primary contributors to the reduction in beam size. While the fact that the first is so useful is not surprising, the utility of the fifth is interesting. We note that the change in the fifth knob required to minimize the beam size was more than twice that of any of the other knobs and more than four times as much as the requisite change in the first knob, suggesting that its importance stems from the fact that our starting lattice was very misaligned in that direction. From Fig. \ref{fig:corr_emit}, we see that the fifth knob also contributes significantly to the reduction of the beam size in simulation when starting from a corrected lattice. Recalling that even our ``uncorrected'' lattice still contains some history of magnet adjustments for CESR's light-source operations, we believe that the reason for the fifth knob's anomalous behavior stems from these prior corrections having already made effective use of the higher-eigenvalue knobs. The fifth knob is then very useful because it is the strongest knob which had not yet been well-tuned. Although we cannot definitively conclude that additional knobs will not give significant further improvements to the beam size, the fact that the fifth knob's exceptional performance is substantiated by simulation, while no other knob shows such striking behavior, suggests that the knobs beyond the first 8 will not act in the same way.

Comparison of the simulated and experimental results shows that, in 
experiment, neither Levenberg-Marquardt nor RCDS brings the vertical 
beam size to its theoretical minimum of a few $\mu$m, obtaining 
instead a lower limit of a few tens of $\mu$m. We infer the existence 
of an unknown source of emittance in the machine that cannot be 
corrected by static magnet changes. A search for the cause had 
previously been made without success \cite{cite:shanks}, and efforts 
in this area are ongoing.

\section{Multi-objective tuning}\label{sec:multi_obj}
\subsection{Theory}

Multi-objective genetic algorithms are useful in myriad accelerator applications since the magnets used to fix one problem will often introduce another; in our case, using vertical correctors as part of an emittance-tuning algorithm introduces orbit errors. When design objectives compete, an intermediate step in 
identifying an optimal solution is to locate the trade-off frontier. 
One design dominates another if it is superior with respect to one 
objective and is not inferior in any of the others. The trade-off 
frontier is the set of non-dominated designs. A multi-objective 
genetic algorithm searches for the trade-off frontier by creating a 
random sampling of initial ``individuals'' in the space of tunable 
parameters, evaluating their merit functions, and preferentially 
breeding the best individuals by combining their tunable parameters, 
or genes \cite{cite:deb}. The population is iteratively grown then 
culled so that surviving individuals converge on the trade-off frontier. 
Genetic algorithms have an essential place in the toolkit of 
accelerator designers
\cite{cite:bazarov,cite:hofler,cite:ehrlichman,cite:borland}, but are 
relatively rare in online applications.

We investigate a multi-objective genetic algorithm that takes two 
merit functions: the vertical beam size at the VBSM and the sum of the 
squares of three vertical orbit measurements in the neighborhood of 
the narrow-aperture undulator. The algorithm is a modification of 
SPEA2 \cite{cite:spea2} as implemented by the PISA collaboration 
\cite{cite:pisa} using a binary tournament selection operator, simulated binary crossover recombination operator, and polynomial mutation operator \cite{cite:deb}. SPEA2 assigns each individual a strength based on 
how many individuals it dominates, then determines the fitness of a 
given individual by summing the strengths of the individuals which 
dominate it, with lower-fitness individuals being better. In order to 
promote diversity, SPEA2 assigns a preference to individuals located 
in more sparsely-populated regions of the objective space. SPEA2 is an 
elitist algorithm, so that both parents and offspring compete for 
inclusion in the next generation.

To handle noisy experimental data, we use sample averaging and modify SPEA2 to randomly resample individuals in the population. As was noted in \cite{cite:huang_rcds1}, in an elitist algorithm, individuals which obtain good merit functions due to noise remain in the population indefinitely. By determining the fitness of an individual by averaging multiple measurements, we reduce the effective noise, and so reduce the probability of such a situation arising in the first place. By periodically resampling individuals, we remove any individual which does end up appearing good solely due to the noise. This was implemented by reevaluating each individual once every seven generations, with the time of the first resampling being random. As was noted in \cite{cite:tian_genetic}, resampling also reduces our sensitivity to machine drifts. Additional parameters used in the genetic algorithm are displayed in Tab. \ref{tab:spea2_parameters}.

\begin{table}[!hbt]
   \centering
   \caption{SPEA2 Parameters.}
   \begin{tabular}{lc}
       \toprule
       Variable Swap Probability & 0\\
       Individual Mutation Probability & 1\\
       Variable Mutation Probability & 0.1\\
       Individual Recombination Probability & 1\\
       Variable Recombination Probability & 1\\
       Eta Mutation & 20\\
       Eta Recombination & 15\\
       \toprule 
   \end{tabular}
   \label{tab:spea2_parameters}
\end{table}

\subsection{Simulation}

Our aim in simulation is to study the speed-up effect that dimension-reduction and the use of conjugate directions provide to the genetic algorithm. We use the values of the 
81 useful corrector magnets as our genes in one set of tests, and 
compare with a second set of tests that use as genes the values of the 
8 eigenparameters, with both run with 30 individuals for 30 generations. The population size was chosen so that we would be able to evaluate one generation of individuals in a reasonable time for online tests. Although this limits the coverage of high-dimensional spaces, we had found in simulation that the genetic algorithm using the corrector magnets as genes performs much better in this configuration than when using a population of 90 with 10 generations, keeping the number of function evaluations constant. We 
simulate the standard CESR lattice with random misalignments without additional 
emittance tuning, iterating over an ensemble of 179 sets of misalignments. For 
each set of misalignments, we combine the final generations from the algorithm using the 81 corrector magnets as genes and the algorithm using the 8 eigenparameters as genes and compute the set of non-dominated individuals. In 114 trials (64\% of trials), the majority of individuals on the joint non-dominated front comes from using the 8 knobs as 
genes, while the reverse occurs in 57 trials (32\% of trials). In this way, we see that using the 8 eigenparameters 
as our genes can improve the performance of the genetic algorithm. We 
expect that, since it has a larger space to explore, the algorithm 
running with the 81 magnets as the genes will eventually obtain a 
superior set of solutions. However, for online problems, speed is a 
more important concern.

Two effects may explain the improved performance of the 8 knobs relative to the 81 magnets. The first cause is the restriction of the search to a lower-dimensional space. The second is the use of conjugate directions. To attempt to differentiate these causes, we use a properly scaled Hadamard matrix to transform the 8 knobs we had used previously into 8 ``mixed'' knobs, such that these mixed knobs are still orthonormal and span the same space, but each is a linear combination of equal parts of our 8 original knobs. By running the genetic algorithm with these knobs, we maintain the advantages of the reduced dimensionality of the problem, but without any advantages which may arise from using conjugate directions. We also run the genetic algorithm with all 81 knobs as genes, so that we maintain the advantages of conjugate directions without dimension-reduction. As a less extreme case, we additionally run a genetic algorithm using the first 16 knobs as genes. We perform tests similar to those described in the preceding paragraph: for the same ensemble of 179 sets of magnet misalignments, we run the genetic algorithm using these 8 mixed knobs, 16 knobs, or 81 knobs as the genes for 30 generations with a population of 30. To compare the relative utility of two sets of genes, for each set of misalignments we combine the final generations of the two algorithms and compute the non-dominated front. We determine how many times the set of individuals found by each algorithm composes the majority of the joint non-dominated front. The results of these tests are shown in Tab. \ref{tab:knob_comparisons_narrow}. We observe that the algorithms using the conjugate-directions as genes routinely outperform the algorithms which do not, regardless of the dimensionality. From this, we conclude that the use of conjugate directions consistently plays a significant role in improving algorithm performance. The dimensionality of the search space does not appear to play a significant role in determining the performance of the genetic algorithm.

\begin{table*}[!hbt]\centering
   \centering
   \caption{Comparison of Genetic Algorithms with Different Genes. The fraction of trials (out of 179) where the individuals found by the genetic algorithm running with the genes listed on the left comprise the majority of the joint non-dominated front when its final population is combined with the final population of the algorithm running with the genes listed above. Higher numbers indicate better performance by the genes to the left and poorer performance by the genes above. In general, the use of conjugate directions significantly improves the convergence of the genetic algorithm.}
   \begin{tabular}{l|c|c|c|c|c}
       & 8 Knobs & 8 Mixed Knobs & 16 Knobs & 81 Knobs & 81 Magnets\\
       \toprule
       8 Knobs & - & 0.63 & 0.46 & 0.51 & 0.64\\
       \hline
       8 Mixed Knobs & 0.34 & - & 0.31 & 0.35 & 0.43\\       
       \hline
       16 Knobs & 0.50 & 0.65 & - & 0.49 & 0.64 \\
       \hline
       81 Knobs & 0.42 & 0.60 & 0.46 & - & 0.53\\
       \hline
       81 Magnets & 0.32 & 0.52 & 0.30 & 0.43 & -\\       
       \toprule 
   \end{tabular}
   \label{tab:knob_comparisons_narrow}
\end{table*}

The fact that the use of conjugate directions improves the performance of the genetic algorithm should not come as a surprise. There is a direct relationship between the value of a knob and the beam size, independent of the values of the other knobs. When using knobs as genes, it then makes sense to talk about one gene having a ``good'' or ``bad'' value and having that reflected in the merit of the individual.

In order to understand the irrelevance of the number of dimensions, consider the example of using 8 versus 81 knobs as our genes. In the latter case, although we have an additional 73 knobs to optimize, they lie along the sloppy directions, and so are not very important to the determination of the emittance. With the bounds chosen for our search space, the increase in emittance due to these lower-eigenvalue knobs is small relative to the changes in emittance from the first few knobs until the latter become very well-optimized, so that even the 81-knob genetic algorithm will be able to cleanly tune the high-eigenvalue knobs. Even if not using knobs as our genes, almost any choice of 8 parameters can be projected onto the space spanned by the first 8 eigenparameters, with the complication that tuning with these parameters entails also altering the other, less-useful 73 eigenparameters. Adding additional parameters only allows the independent tuning of these 73 eigenparameters, the values of which do not have a large impact on the emittance.

If we attempt tuning with the allowed ranges of the knobs and magnets increased by a factor of 10 above what we had used previously, we find that the dimensionality of the search becomes the dominant factor for the convergence of the search. This may be seen in Tab. \ref{tab:knob_comparisons_wide}, which shows the results of making the same comparisons as were performed above, combining the final populations of algorithms run with different genes and seeing how many times each algorithm finds the majority of the individuals on the non-dominated front. The reason for this is that, in the high-dimensional cases, the lower-eigenvalue knobs are allowed to vary enough that they become significant relative to the first few knobs well before the latter are well-tuned. The algorithm then needs to spend resources to optimize these additional knobs as well. When using a reduced number of knobs, the default values for the lower-eigenvalue knobs are fixed near their optimal values relative to the size of the search space, so that, for our chosen number of generations, their unavoidable contribution to increasing the emittance is small relative to the contributions from the high-eigenvalue knobs.

\begin{table*}[!hbt]\centering
   \centering
   \caption{Comparison of Genetic Algorithms with Different Genes in a Larger Search Space. The fraction of trials (out of 179) where the individuals found by the genetic algorithm running with the genes listed on the left comprise the majority of the joint non-dominated front when its final population is combined with the final population of the algorithm running with the genes listed above. Higher numbers indicate better performance by the genes to the left and poorer performance by the genes above. The search range for each gene has been increased by a factor of 10 from what was used to obtain Tab. \ref{tab:knob_comparisons_narrow}. In this case, the reduction in the number of dimensions plays a significant role in improving the algorithm performance.}
   \begin{tabular}{l|c|c|c|c|c}
       & 8 Knobs & 8 Mixed Knobs & 16 Knobs & 81 Knobs & 81 Magnets\\
       \toprule
       8 Knobs & - & 0.54 & 0.57 & 0.79 & 0.99\\
       \hline
       8 Mixed Knobs & 0.42 & - & 0.56 & 0.74 & 0.99\\       
       \hline
       16 Knobs & 0.37 & 0.39 & - & 0.61 & 0.96 \\
       \hline
       81 Knobs & 0.19 & 0.23 & 0.36 & - & 0.96\\
       \hline
       81 Magnets & 0.00 & 0.01 & 0.03 & 0.02 & -\\       
       \toprule 
   \end{tabular}
   \label{tab:knob_comparisons_wide}
\end{table*}

\subsection{Experiment and Discussion}

\begin{figure}[!htbp]
\begin{center}
\includegraphics[width=1.0\columnwidth]{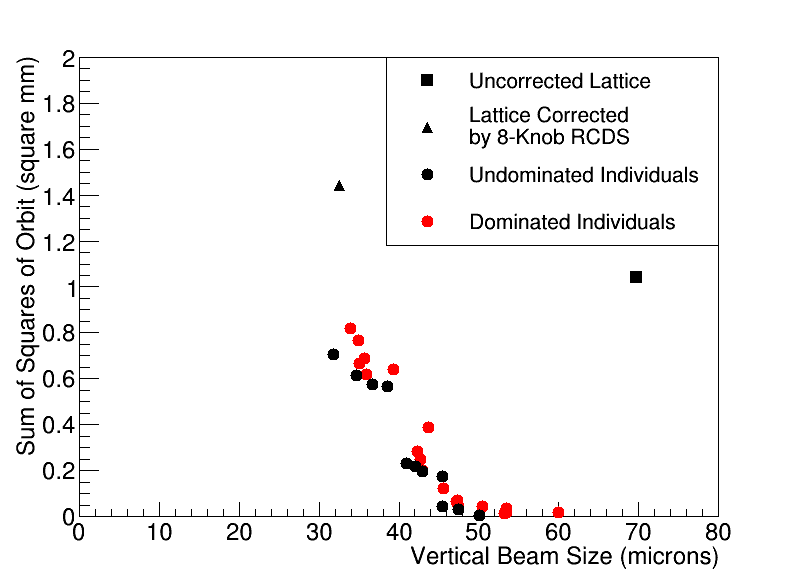}
\end{center}
\caption{\label{fig:real_genetic_1} The final generation of our 
genetic algorithm as applied to CESR, with the non-dominated 
individuals differentiated from the rest. Also plotted are the initial 
beam size and orbit and those obtained after the 8-knob RCDS tuning, 
as in Fig. \ref{fig:start_uncorrected}. The BPMs near the undulator had some offsets, so we are in fact attempting to steer the beam onto an arbitrary off-axis trajectory. Given that steering the beam onto a specific non-zero orbit also represents a trade-off with respect to minimizing emittance, our conclusions are not affected.}
\end{figure}

Running the genetic algorithm on CESR, we use the leading 8 
eigenparameters as the genes and initialize the machine with the 
lattice and conditions for light-source operations, without additional 
emittance tuning. The initial vertical beam size is 70 $\mu$m. After 
running with a population of 30 individuals for 11 generations, we 
obtain beam sizes of 30 $\mu$m, as shown in Fig. 
\ref{fig:real_genetic_1}. This performance is comparable to the RCDS 
algorithm, but with greater control over the orbit. With the 
population size chosen, each generation may be evaluated in roughly 
ten minutes.

\begin{figure}[!htbp]
\begin{center}
\includegraphics[width=1.0\columnwidth]{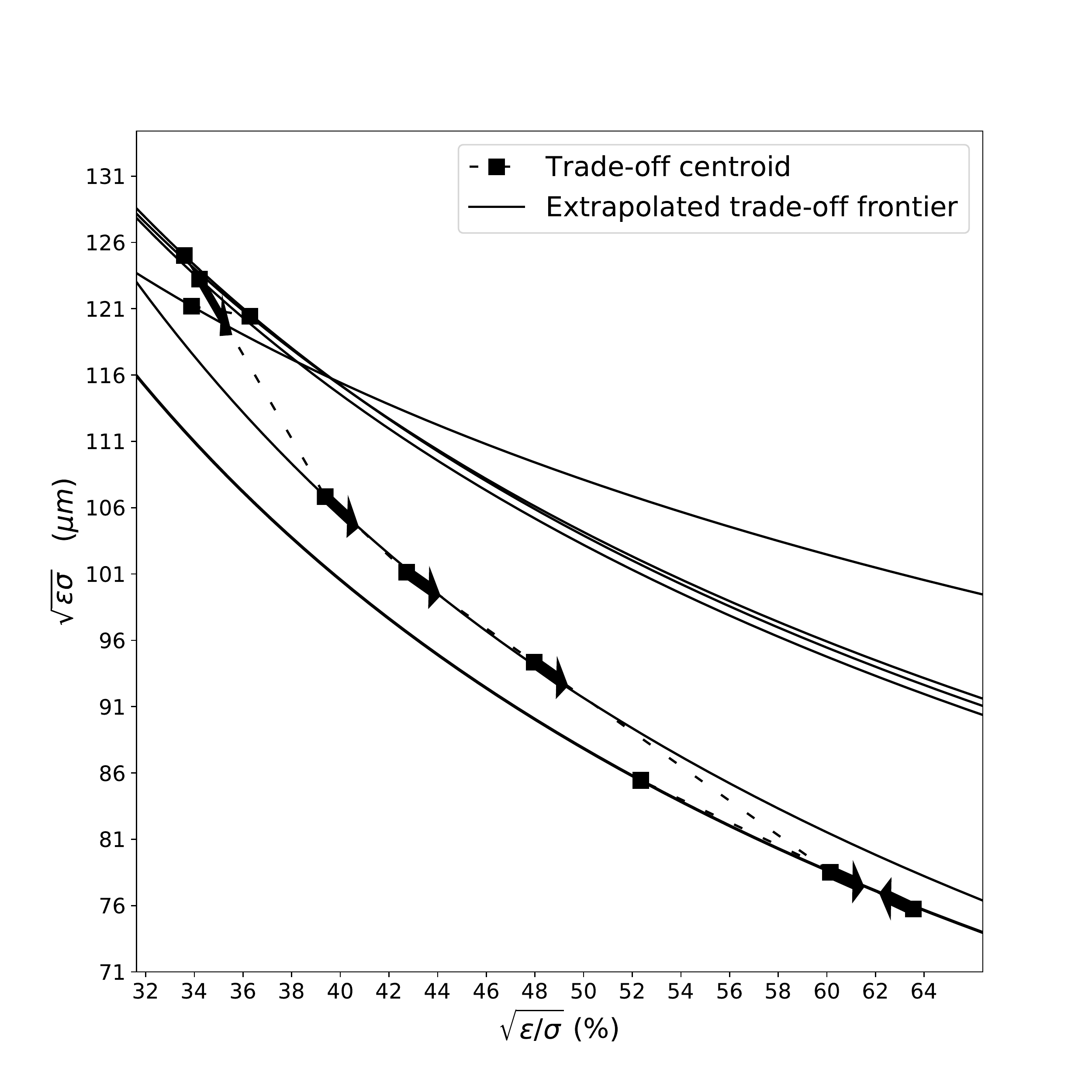}
\end{center}
\caption{\label{fig:real_genetic} Experimental progress of the genetic 
algorithm with time. Each square marks a generation of the algorithm, 
showing the location of the mean individual in the trade-off frontier. 
Arrows indicate the sequence of generations, with one generation 
taking 10 minutes of machine time. The $x$ axis shows the square root 
of the ratio of the RMS beam size $\epsilon$ and RMS orbit error 
$\sigma$, while the $y$ axis shows the square root of their product. 
For any given $x$, smaller $y$ values are better. 
Solid lines show the best power-law fit $(y=Ax^b)$ to the trade-off 
frontier at each generation. Much of the improvement provided by the genetic algorithm is due to reduction of the orbit error, as evidenced by the trend of the algorithm toward larger values of $\sqrt{\epsilon/\sigma}$.}
\end{figure}

\begin{figure}[!htbp]
\begin{center}
\includegraphics[width=1.0\columnwidth]{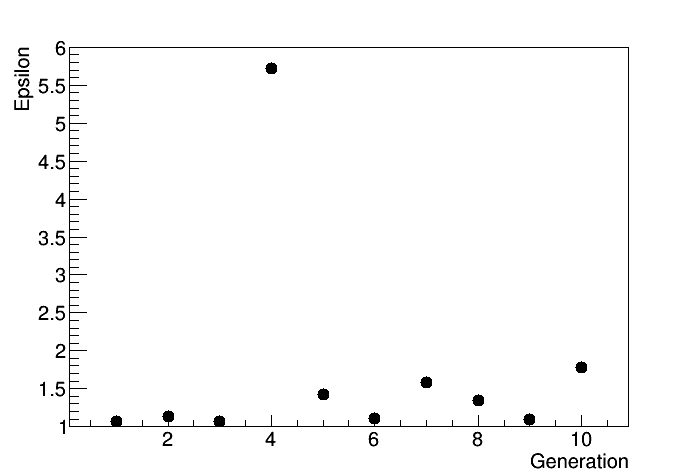}
\end{center}
\caption{\label{fig:real_convergence} The {\em epsilon} test of 
convergence in our online study: $\epsilon$ is the scale factor such 
that the $i^{th}$ generation frontier set $X_i$ dominates the rescaled 
frontier set $\epsilon X_{i+1}$. The figure's vertical axis shows 
$\epsilon - 1$. The algorithm converges as $\epsilon$ approaches 
unity. The figure shows that the algorithm is still providing improvement at the end of the allotted time. Comparison with Fig. \ref{fig:real_genetic} shows that this improvement is mainly due to reduction in the orbit error.}
\end{figure}

It is also interesting to measure the rate of convergence to the trade-off frontier. The progress of our algorithm is shown in Fig.~\ref{fig:real_genetic}. There is no consensus in the computer science literature on the best way to measure multi-objective convergence \cite{riquelme2015performance}. We choose the most popular metric among the subset that do not assume knowledge of the ``true'' trade-off frontier, the {\em epsilon test} \cite{zitzler2003performance}. Given the trade-off frontiers $X_i$ and $X_{i+1}$ at iterations $i$ and $i+1$ of the algorithm, $\epsilon$ is defined as the minimum scaling factor such that every element of the rescaled $\epsilon X_{i+1}$ is dominated by at least one element of $X_i$. The algorithm converges as $\epsilon\rightarrow1$, meaning that successive generations just barely dominate their predecessors and the fitness of the gene pool is not improving over time. We find running on the real machine that the algorithm  has not fully converged within the time allotted for it to run, as is shown in Fig.~\ref{fig:real_convergence}. This is consistent with the findings from \cite{cite:huang_genetic,cite:tian_genetic} that genetic algorithms converge slowly when applied to online optimization. However, we emphasize that the genetic algorithm need not converge to be useful; good intermediate solutions found by a genetic algorithm, such as those shown in Fig. \ref{fig:real_genetic_1},; still represent good working points for the machine.

\section{Conclusions and Outlook}\label{sec:conclusion}

We have shown that, by making the proper choice of decision variables, 
we are able to reduce the 81-dimensional task of tuning the vertical 
emittance at CESR to an 8-dimensional problem with little loss in our 
ability to minimize the beam size. These few stiff knobs enable the 
efficient use of the RCDS algorithm for tuning the machine. We have 
also demonstrated that stiff knobs speed up the convergence of a 
multi-objective genetic algorithm. The utility of the genetic 
algorithm extends to online optimization of the real machine.

It is important to note that with either the genetic or RCDS algorithms, using more knobs will enable one to find an improved solution, but at the cost of increased time of running. For our tests, the fact that we have repeatedly demonstrated that 10\% of the available knobs (8 out of 81) are able to provide at least 50\% of the potential improvement in beam size (and much more if there was some additional source of emittance in our machine which could not be corrected by such knobs) shows that this choice is a useful compromise between speed of execution and utility of results. Time and performance constraints will inform the optimal choice of the number of knobs for practical use elsewhere.

In addition to the cases reported here, this dimension-reduction 
method will enable other algorithms that scale poorly with the number 
of free parameters to be applied to accelerators. Moreover, the fact that accelerators appear to display features of sloppy systems motivates the application of techniques for simplifying nonlinear sloppy systems to problems such as injection or lifetime optimization. We also aim to apply this work to the tuning of other high-dimensionality systems, such as 
electron microscopes. To facilitate the above objectives, we are 
planning to create a more universal toolkit for flexibly applying 
various algorithms to a myriad of accelerator systems based on an 
Experimental Physics and Industrial Control System (EPICS) interface.
It will also be interesting to explore ways to correct for the fact that prior tuning reduces the expected effectiveness of some of the high-eigenvalue knobs.

\newpage

\section{Acknowledgements}

We would like to thank Jim Shanks for assistance with the simulations, 
Suntao Wang for managing the VBSM, He He for computations of the 
Hessian matrix, and the CESR operators and support staff for ensuring 
the smooth running of these experiments. We would like to acknowledge 
the support of the US Department of Energy through grant number DE-SC 
0013571. W.F.B. would like to acknowledge the support of the 
National Science Foundation Graduate Research Fellowship Program under 
grant number DGE-1650441. C.J.R.D., D.B.L., I.V.B., and J.P.S. were also partially supported by the US National Science Foundation under Award OIA-1549132, the Center for Bright Beams.

\bibliography{let_knobs_paper}

\end{document}